\documentclass[lettersize,journal]{IEEEtran}
\usepackage{amsmath,amsfonts}
\usepackage{amsthm}
\usepackage{algorithmic}
\usepackage{algorithm}
\usepackage{array}
\usepackage[caption=false,font=normalsize,labelfont=sf,textfont=sf]{subfig}
\usepackage{textcomp}
\usepackage{stfloats}
\usepackage{url}
\usepackage{verbatim}
\usepackage{graphicx}
\usepackage{cite}
\usepackage[hidelinks]{hyperref}
\usepackage{orcidlink}
\usepackage[capitalise]{cleveref}
\usepackage{bbold}
\usepackage{braket}
\usepackage{dsfont}   
\usepackage{graphicx}
\usepackage{dcolumn}
\usepackage{bm}
\usepackage[T1]{fontenc}
\usepackage[dvipsnames]{xcolor}
\usepackage{comment}
\usepackage{tikz}
\usepackage{quantikz}
\usepackage{ragged2e}
\usepackage{rotating}
\usepackage{braket}
\usepackage{adjustbox}
\usepackage{todonotes}
\usepackage{tcolorbox}

\usetikzlibrary{tikzmark}

\DeclareMathOperator{\Tr}{Tr} 

\usepackage{ulem}

\definecolor{mintgreen}{RGB}{152, 255, 152}

\usepackage{comment}

\begin{document}

\title{Controlled Continuous-Time Quantum Walks for Power Grid Analysis under N-x Contingencies}

\author{
Mahesh~N.~Jayakody\,\orcidlink{0000-0002-1486-6097},
Ruben Kara\,\orcidlink{0000-0001-9258-4397},
Carsten Hartmann\,\orcidlink{0009-0007-5067-589X},
and Andrea Benigni\,\orcidlink{0000-0002-2475-7003}%
\thanks{All authors are with the Institute of Climate and Energy Systems (ICE-1), Forschungszentrum Jülich, 52428 Jülich, Germany
(e-mail: m.jayakody@fz-juelich.de;
r.kara@fz-juelich.de;
c.hartmann@fz-juelich.de;
a.benigni@fz-juelich.de).}%
}



\maketitle

\begin{abstract}
The combinatorial growth of network configurations under $N-x$ contingencies presents significant computational challenges for power system analysis. In this work, we introduce a quantum algorithm based on Controlled Continuous-Time Quantum Walks (CCTQWs) that evolves a quantum walker conditioned on a contingency register held in a uniform superposition of all $x$-line outage configurations representing the contingency scenarios. The proposed framework encodes transmission line availability in a quantum register and employs controlled Hamiltonian evolution to account for topology changes induced by contingencies. We develop a gate-based implementation using Trotter–Suzuki product formulas and investigate the associated approximation errors and computational requirements. As a demonstration of the proposed methodology, we introduce a contingency-averaged occupation centrality measure derived from quantum-walk measurement statistics. Numerical experiments on an 8-bus power network under $N-2$ contingencies demonstrate the implementation of the algorithm and characterize the convergence of the resulting centrality estimates with respect to the observation window, sampling and Trotter steps. Comparison with established classical centrality measures shows a qualitatively consistent characterization of network structure, particularly in identifying the most central nodes. The proposed framework establishes a methodological foundation for quantum-walk-based analysis of power networks under contingencies, providing a starting point for future investigations of quantum algorithms for power system security assessment.
\end{abstract}


\section{Introduction}
The number of configurations in $N-x$ contingency analysis grows combinatorially with power system size. An increase in $N$ is expected, driven by the need to analyse larger areas beyond transmission systems, including sub-transmission and distribution grids. At the same time, the growing penetration of distributed energy resources significantly raises the number of critical units in the system, leading to scenarios with $x>2$. Consequently, as both $N$ and $x$ increase, exhaustive contingency analysis is expected to become computationally impractical. For example, the ENTSO-E report on the South-East European blackout of 21 June 2024~\cite{ENTSOE2025GridIncident} highlights that, had a larger area and $N-3$ contingency scenarios been considered, the critical conditions leading to the blackout could have been identified prior to the event. At the same time, in a system with 1,000 components, a full $N-3$ analysis already requires about $10^8$ contingencies, rendering the problem practically infeasible.

In recent years, the envisioned growth in computational cost associated with contingency analysis has motivated the use of topological graph--based measures, such as centrality measures (CMs), to prioritize components that are structurally important under multiple outage scenarios, and thus guiding the contingency analysis \cite{halappanavar2012towards, coelho2018new,srivastava2018graph,narimani2021generalized}. One of the main advantages of employing CMs is their ability to analyze the grid system with incomplete information \cite{srivastava2018graph}. 

Motivated by this, we aim to develop a Quantum Walk (QW)-based framework for power-grid analysis under $N-x$ contingencies. We then use this framework to calculate node centrality measures and examine whether the results are qualitatively consistent with established classical centrality measures under $N-x$ contingencies. To the best of our knowledge, the application of QWs to power-grid analysis remains largely unexplored.

A QW is considered the quantum analog of a classical random walk. Similar to its classical counterpart, QWs have been extensively utilized in quantum algorithm development \cite{kempe2003quantum,shenvi2003quantum,kadian2021quantum} and quantum computing \cite{childs2009universal} in general. There are two broad classes of QWs: discrete-time QWs and continuous-time QWs (CTQWs), each of which has significant distinctions in its mathematical formalism \cite{wang2013physical}. Several studies can be found in the literature where CTQWs are employed to study centrality measures in various types of networks. In \cite{boito2023ranking}, four new centrality measures for directed networks are introduced based on CTQWs. In \cite{rossi2014node}, the vertex centrality, a measure that quantifies the relative importance of a node within a network, is linked to the interference patterns of the CTQWs by defining a new centrality measure using the Quantum Jensen-Shannon Divergence. Moreover, the ability of CTQWs to rank important nodes on large-scale weighted graphs is shown in \cite{wang2022continuous}. 

While QWs have been previously employed to define centrality measures on single graphs, we develop a Controlled Continuous-Time Quantum Walk (CCTQW) framework that enables coherent QW evolution over a superposition of all $\binom{N}{x}$ contingency-induced subgraphs, where $\binom{N}{x}$ is the binomial coefficient. The walk evolution depends explicitly on the operational status of transmission lines. From the resulting measurement statistics, we derive a contingency-averaged occupation centrality measure as an application of the proposed framework. The main contributions of this work are summarized as follows

\begin{itemize}
     \item We develop a CCTQW framework that extends QW evolution from a single graph to a superposition of $\binom{N}{x}$ contingency-induced subgraphs, with the evolution conditioned on the status of transmission lines.
    \item We provide a gate-based implementation of proposed framework using Trotter--Suzuki product formulas and discuss approximation errors and mitigation strategies.
    \item We define a node-ranking metric from measurement statistics, capturing contingency-averaged importance.
    \item We demonstrate the proposed approach on an 8-bus test system for $N-2$ contingencies and compare the results with classical centrality measures.
\end{itemize}

The remainder of this paper is organized as follows. \cref{sec:1d} introduces the background and mathematical framework, including the graph representation of power grids and the CTQW formulation. \cref{Contingency-Controlled-CTQW} presents the proposed CCTQW algorithm and the contingency-aware centrality measure. \cref{detailed_implementation} provides the detailed gate-level circuit implementation. \cref{Trotter Correction} discusses Trotter--Suzuki approximation errors and mitigation strategies. \cref{Particular Case} reports numerical results on an 8-bus test system, including comparison with classical centrality measures. The paper concludes with a discussion on resource requirements, limitations, and possible future directions, followed by concluding remarks.

\section{Background and Mathematical Framework} \label{sec:1d}
This section introduces the graph-theoretic representation of power grids, the mathematical formulation of CTQWs, and their implementation on gate-based quantum hardware using Trotter-Suzuki product formulas. These elements provide the foundation for the contingency-controlled evolution introduced in \cref{Contingency-Controlled-CTQW}.

\subsection{Graph representation of Power grids} 
An electric power system can be represented as a finite undirected graph $G=\{V,E\}$ where $V=\{1 , \hdots, n\}$ denotes the set of $n$ nodes, corresponding to buses or substations in the power grid. The edge set $E=\{(r,s): r,s \in V\}$ represents the connections between buses, e.g. transmission lines. The number of lines is denoted by $N=|E|$. For convenience, we enumerate the edges as $e_{r,s}=1,\dots,N$, where each edge $e_{r,s}$ corresponds to the node pair $(r,s)\in E$. Throughout this paper, the terms \emph{nodes} and \emph{buses} are used interchangeably, as are \emph{edges} and \emph{transmission lines}.

\subsection{CTQW on a graph} 
A CTQW on a graph $G$ with $n$ vertices describes the unitary time evolution of a quantum walker whose state $|\psi(t)\rangle$ lies in the Hilbert space $\mathcal{H}_w$ spanned by the computational basis states $V_w = \{|1\rangle, \hdots, |n\rangle\}$, where each basis state $|j\rangle$ corresponds to vertex $j$ of $G$. The evolution of the CTQW on the graph $G$ is governed by the Schr\"odinger equation ($\hbar = 1)$
\begin{equation}
\label{eq:schrödinger}
    i \frac{d}{dt}|\psi(t)\rangle = H|\psi(t)\rangle,
\end{equation}
where $H$ is the Hamiltonian of the system. Typically, $H$ is chosen to be either the adjacency matrix or the Laplacian of the underlying graph. While the Laplacian may seem more natural from a power systems perspective, given its connection to the nodal admittance matrix, the adjacency matrix has a purely off-diagonal structure, which simplifies the gate decomposition of the edge-local evolution operators. Hence, in this work, we adopt the adjacency matrix $\mathcal{A}$ of $G$ directly as the Hamiltonian that governs the dynamics of the walker. The (weighted) adjacency matrix is defined as 
\begin{equation}
    \mathcal{A}_{r,s} =
\begin{cases}
w_{r,s} & \text{if } (r,s) \in E \\
0 & \text{otherwise}
\end{cases},
\end{equation}
where $w_{r,s} > 0$. The adjacency matrix representation naturally distinguishes between two types of graphs. When all edge weights are unity (i.e., $w_{r,s}=1$ for all $(r,s) \in E$), the graph is termed \emph{unweighted}. Otherwise, it is referred to as a \emph{weighted} graph. In the context of power networks, one can use an unweighted graph to represent only the connectivity structure of the grid, i.e., whether two buses are connected by a transmission line. A weighted graph, on the other hand, can be used to model the power grid by associating a physical property of the transmission lines. For instance, the weights may represent quantities such as line strength or the magnitude of line admittance. When only geographic information is known, the weights can be used to represent the inverse of the line length.

By solving the Schr\"odinger equation, the quantum walker's state is obtained as $|\psi(t)\rangle=e^{-itH}|\psi(0)\rangle
=\sum_{j=1}^{n}\alpha_j(t)|j\rangle$,
where $\alpha_j(t)\in\mathbb{C}$. The probability of finding the walker at node $j$ at time $t$ is
$P_j(t)=|\alpha_j(t)|^2$. The corresponding occupation (time-averaged) centrality is defined as \cite{boito2023ranking,wang2022continuous}

\begin{equation}\label{centrality_measure}
    \tilde{C}_j= \lim_{T \to \infty} \frac{1}{T} \int_{0}^{T} P_j (t) dt.
\end{equation}

In practice, the asymptotic time average in \cref{centrality_measure} is estimated over a finite observation horizon $T$ by sampling the occupation probabilities at discrete times $t_{\tau}$. The occupation (time-averaged) centrality is then approximated as

\begin{equation}
\label{centrality_discrete}
\tilde{C}_{j,\mathrm{approx}} = \frac{1}{N_T}\sum_{\tau=1}^{N_T} P_j(t_{\tau}),
\end{equation}
where $N_T$ is the number of samples, $t_{\tau}= \tau\Delta t$ for $\tau=1,\dots,N_T$ and $\Delta t=T/N_T$. The accuracy of this approximation depends on the finite observation horizon $T$ and the sampling interval $\Delta t$. 

\subsection{Implementation of CTQW on a gate-based architecture}

From a computational perspective, evaluating the time-averaged occupation probability $\tilde{C}_{j,\mathrm{approx}}$ requires implementing the unitary evolution operator $e^{-itH}$ in terms of basic quantum gates. Several approaches have been proposed for realizing CTQWs on gate-based quantum computers, with efficient circuit constructions exploiting graph symmetry or sparsity \cite{chen2024implementation, loke2017efficient, portugal2025efficient}. For a CTQW on a general graph, the evolution operator is typically implemented using Hamiltonian-simulation techniques. When $H$ is expressed as a sum of local terms, a standard strategy is to approximate the evolution using product-formula (Trotter--Suzuki) decomposition \cite{childs2019faster, zhang2023low, campbell2019random}. 

We now discuss the implementation of the CTQW via Hamiltonian simulation for a single graph topology, since it serves as the fundamental building block of the proposed framework. Since the Hamiltonian is set as the adjacency matrix, it can be decomposed into a sum of edge-local transition matrices $H^{(r,s)}$, each associated with an edge $(r,s)$, so that $H=\sum_{(r,s)}H^{(r,s)}$. We define a matrix $M^{(r,s)}$ of the following form $M^{(r,s)}=(v_r v_s^\text{T} + v_s v_r^\text{T})$ where $v_k \in \mathbb{R}^{n}$ is the $k$-th standard basis column vector with a $1$ in the $k$-th position and $0$ elsewhere. Now, we define the edge--local transition matrix as
\begin{equation}\label{decom_H}
    H^{(r,s)} = w_{r,s} M^{(r,s)},
\end{equation}
where $w_{r,s}$ is the $(r,s)$-th entry of the adjacency matrix. Note that, for a general dense graph, such an edge-wise decomposition may lead to a large number of local Hamiltonian terms, resulting in an inefficient implementation when using product-formula methods. However, power networks are inherently sparse. This feature enables the decomposition of the edges into commuting sets to efficiently reduce discretization errors (see \cref{sec:sub_regrouping}). As the local transition matrices do not generally commute i.e. $[ H^{(r,s)}, H^{(r^\prime,s^\prime)}] \neq 0$, using Lie product formula yields
\begin{equation}
    e^{-itH} = \lim_{R\rightarrow\infty}
    \left[\prod_{(r,s)} e^{-i\frac{t}{R} H^{(r,s)}}\right]^R. 
\end{equation}
Using finite steps $R$, we can approximate the time evolution according to 
   
\begin{subequations}
\begin{align}
e^{-itH} &\approx \left[\prod_{(r,s)} e^{-i\frac{t}{R} H^{(r,s)}}\right]^R
\label{eq:single_qw_a}\\[6pt]
e^{-itH} &\approx
\left[
\overset{\rightarrow}{\prod}_{(r,s)}
e^{-i \frac{t}{2R}H^{(r,s)}}
\overset{\leftarrow}{\prod}_{(r,s)} 
e^{-i \frac{t}{2R}H^{(r,s)}}
\right]^R \label{eq:single_qw_b}
\end{align}
\end{subequations}
where the arrow $\rightarrow$ indicates that the product over edges follows a fixed ordering of $(r,s)$, while $\leftarrow$ denotes the reverse ordering. Equation \cref{eq:single_qw_a} is called first-order Lie–Trotter formula and \cref{eq:single_qw_b} corresponds to the second-order symmetric (Strang–Suzuki) formula. To improve the accuracy–cost tradeoff, higher-order product formulas may be employed to achieve the same simulation accuracy with fewer Trotter steps. Further details can be found in \cite{childs2019faster}. The product formulas in \cref{eq:single_qw_a} and \cref{eq:single_qw_b} can be implemented on a gate-based quantum computer by decomposing each term $e^{-i\frac{t}{R} H^{(r,s)}}$ or $e^{-i \frac{t}{2R}H^{(r,s)}}$ into a sequence of single- and multi-qubit gates. In practice, $R$ is kept finite, which introduces an approximation error that depends on the number of steps $R$, the total evolution time $t$, and the spectral properties of the Hamiltonian. 

\section{Contingency-Controlled-CTQW} \label{Contingency-Controlled-CTQW}
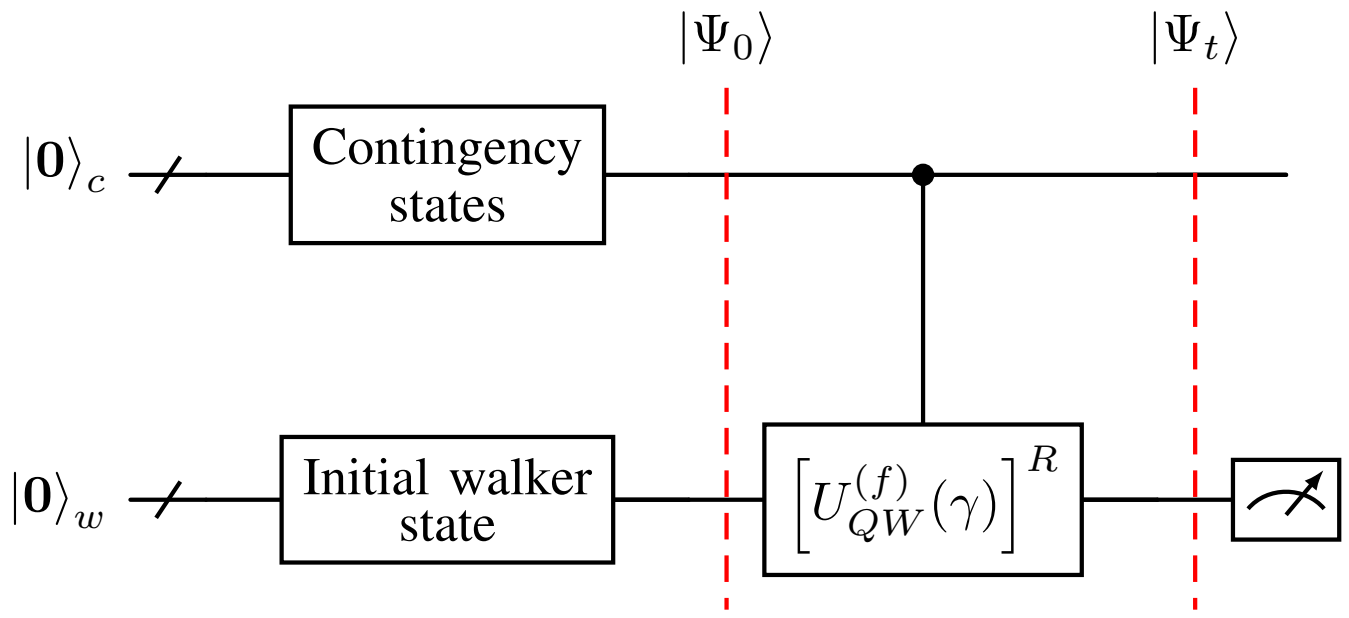
\begin{figure}[t]
\centering
\begin{quantikz}[row sep=1.2cm]
\lstick[label style={align=center}]{\shortstack{$\ket{\mathbf{0}}_c$\\[0.2ex]}}
  & \qwbundle{} 
  & \gate{\shortstack{Contingency\\states}} 
  & \slice{$\ket{\Psi_0}$}
  & \ctrl{1} 
  & \slice{$\ket{\Psi_t}$}
  & \qw \\
\lstick[label style={align=center}]{\shortstack{$\ket{\mathbf{0}}_w$}}
  & \qwbundle{} 
  & \gate{\shortstack{Initial walker\\state}}
  & \qw
  & \gate{\left[U_{QW}^{(f)}(\gamma)\right]^R} 
  & \qw
  & \meter{}
\end{quantikz}
\caption{Conceptual quantum circuit for the CCTQW. The contingency ($N$ qubits) and walker ($\lceil \log n\rceil$ qubits) registers are prepared into the joint state $|\Psi_0\rangle$ and evolved under $\left[U_{QW}^{(f)}(\gamma)\right]^R$ to yield  $|\Psi_t\rangle$, from which the sampling is performed. Here, $f$ denotes the order of Trotter-Suzuki formula used, $\gamma \in \mathbb{R}$ is a tunable parameter and $R$ is the number of Trotter steps.}
\label{circuit1}
\end{figure}

In this section, we generalize the time evolution of the quantum walker along one single graph to a superposition over $\binom{N}{x}$ configurations, each encoded as a computational basis state of a contingency register and evolved under an edge-controlled evolution. The overall procedure is conceptually summarized in \cref{circuit1}. The circuit in \cref{circuit1} comprises of two registers: The contingency register (top quantum register) is used to encode topologies of all $\binom{N}{x}$ subgraphs. It consists of $N$ qubits, where $N$ represents the number of edges in the power grid. The n-th qubit in the contingency register encodes whether the n-th edge is active (1) or inactive (0). The walker register (bottom quantum register) encodes the nodes of the power grid on which the quantum walker propagates. It consists of $\lceil \log n \rceil$ qubits, where $n$ is the number of nodes and $\lceil \cdot \rceil$ denotes the ceiling function. First, the contingency register is prepared in an equal superposition of all configurations corresponding to $x$ simultaneous line outages, while the walker register is initialized in a uniform superposition over the network nodes. The joint state is then evolved under a controlled CTQW, such that each basis state of the contingency register determines the graph topology governing the walker dynamics. Finally, the walker register is sampled at selected evolution times, and the resulting occupation probabilities are used to construct a contingency-aware node-ranking measure. The following subsections describe the main components of the CCTQW framework, while the detailed gate-level implementation is presented in \cref{detailed_implementation}.

\subsection{State preparation}
\label{sec:state_prep}
The preparation of the contingency states and the initial walker state is a crucial component of the proposed framework. Since we are interested in all $\binom{N}{x}$ subgraphs, we need to create the superposition of the encoded subgraphs in the contingency register where $x$ qubits are $0$. To address this challenge we use short-depth quantum circuits that deterministically prepare Dicke state $\ket{D^N_x}$, an equal-amplitude superposition of all $N$-qubit computational basis states with a fixed Hamming weight $x$ \cite{bartschi2022short, dicke1954coherence}. Hence, the contingency states are prepared as 
\begin{align}\label{dicke state}
    \ket{D_x^N}=\frac{1}{\sqrt{\binom{N}{x}}}  \sum_{d \in \mathcal{C}^{(x)}} \ket{d}
\end{align}
where $\mathcal{C}^{(x)}$ is the set of all $N$-bitstrings with Hamming weight $x$. For example, with $N=4$ and $x=2$ the corresponding Dicke state reads $\ket{D_2^4}=\frac{1}{\sqrt{6}} ( \ket{1100} + \ket{1010}+ \ket{1001} \nonumber\\ + \ket{0110}+ \ket{0101} + \ket{0011} )$. Note that, since the standard Dicke state $\ket{D_x^N}$ contains $x$ ones and $N-x$ zeros, we apply Pauli-$X$ gates to all qubits after state preparation written as $\ket{\tilde{D}^N_x}$ to obtain a superposition containing $x$ zeros and $N-x$ ones, consistent with the contingency encoding adopted in this work. 

The walker register is initialized in a uniform superposition over all nodes to ensure an unbiased starting point for the dynamics. The initial walker state is given by
\begin{equation}\label{initial_walker}
    |\phi_0\rangle = \frac{1}{\sqrt{n}} \sum_{j=1}^{n} |j\rangle
\end{equation}
where $n$ is the number of nodes. If $n$ is a power of two (i.e., $n=2^m$ for some $m$), the walker register can be initialized using Hadamard gates as $|\phi_0\rangle= H^{\otimes m} |0\rangle^{\otimes m}$. For $n \neq 2^m$, the walker register is embedded into a Hilbert space of dimension $2^{\lceil \log n \rceil}$. In this case, one may prepare a uniform superposition over all $2^{\lceil \log n \rceil}$ basis states using Hadamard gates and restrict the dynamics to the first $n$ states.

Alternatively, one may directly prepare the uniform superposition over the first $n$ basis states using standard quantum state preparation techniques \cite{mottonen2004transformation,soklakov2006efficient} although at the cost of increased circuit depth. The initial state of the algorithm is thus given by the tensor product 
\begin{align}
     \ket{\Psi_0} = \ket{\tilde{D}_x^N} \otimes \ket{\phi_0}. 
     \label{eq:overall_state}
 \end{align}
This joint state serves as the input to the controlled CTQW introduced in the next subsection.

\subsection{Controlled-CTQW on a graph}
To enable edge controlled evolutions, we extend the overall Hilbert space to $\mathcal{H}_c \otimes \mathcal{H}_w$, where $\mathcal{H}_c$ is of dimension $2^N$ and populated by quantum states of the form $\ket{d_1 d_2 \dots d_N}$ where $d_i \in \{0;1\}$ indicating the operational status of edge $i$ i.e. $\ket{1}$ and $\ket{0}$ denote an active and inactive edge, respectively. We call the corresponding quantum register contingency register.

Having extended the Hilbert space, we now introduce a controlled operation whose action on the walker register is conditioned on the state of the contingency register, so that each edge contributes to the walk dynamics only when its associated qubit marks the line as active. For each edge $(r,s)$, let
\begin{equation}
\begin{split}
\Pi_0^{(e_{r,s})} &=
\mathbb{1}^{\otimes(e_{r,s}-1)}
\otimes \ket{0}\bra{0}_{e_{r,s}}
\otimes \mathbb{1}^{\otimes(N-e_{r,s})}, \\
\Pi_1^{(e_{r,s})} &=
\mathbb{1}^{\otimes(e_{r,s}-1)}
\otimes \ket{1}\bra{1}_{e_{r,s}}
\otimes \mathbb{1}^{\otimes(N-e_{r,s})},
\end{split}
\end{equation}
where $\Pi_0^{(e_{r,s})}$ and $\Pi_1^{(e_{r,s})}$ denote the projectors onto the subspaces in which the qubit associated with edge $(r,s)$ is in state $\ket{0}$ and $\ket{1}$, respectively, and $e_{r,s}$ denotes the index of that edge in the enumeration $e_{r,s}\in\{1,\dots,N\}$. Then, the overall Hamiltonian can be written as $ H_{\mathrm{ctrl}}=\sum_{e_{r,s}} \Pi_1^{(e_{r,s})} \otimes H^{(r,s)}$ which activates the contribution of edge $(r,s)$ only when the $e_{r,s}$-th qubit is in state $\ket{1}$. Using Lie product formula, the corresponding overall time-evolution operator is given by

\begin{equation}\label{controlled_qw_op}
e^{-itH_{\mathrm{ctrl}}}
= \lim_{R\rightarrow\infty}
\left[
\prod_{(r,s)}
e^{-i \frac{t}{R}\left(\Pi_1^{(e_{r,s})}\otimes H^{(r,s)}\right)}
\right]^R
\end{equation}
The first- and second-order product-formula approximations of the controlled time-evolution operator for a finite number of Trotter steps $R$ can be obtained by following the same procedure as in \cref{eq:single_qw_a} and \cref{eq:single_qw_b}, respectively.

\subsection{Implementation of CCTQW }

The CCTQW is generated by a unitary operator whose action on the walker register is conditioned on the configurations encoded in the contingency register. Following the controlled time-evolution formulation in \cref{controlled_qw_op}, we define the edge-wise controlled unitary operator as

\begin{equation}\label{eq:Ue_projector_exp}
U^{(r,s)}(\gamma)
= \exp\!\left(-i\,\gamma\, \Pi_1^{(e_{r,s})} \otimes H^{(r,s)}\right)
\end{equation}
where $\gamma \in \mathbb{R}$ is a tunable parameter. The operator $U^{(r,s)}(\gamma)$ governs the transition of the quantum walker between nodes $r$ and $s$, conditioned on the state of the qubit $e_{r,s}$ in the contingency register. Using the projector property, the operator $U^{(r,s)}(\gamma)$ can be written in the explicit form
\begin{equation}\label{Urs}
U^{(r,s)}(\gamma)
= \Pi_0^{(e_{r,s})} \otimes \mathbb{I}
+ \Pi_1^{(e_{r,s})} \otimes e^{-i \gamma H^{(r,s)}}
\end{equation}
which shows that the evolution generated by $H^{(r,s)}$ is applied only when the corresponding contingency qubit is in the state $\ket{1}$, and otherwise the identity operation is applied. Finally, the CCTQW operator $U_{QW}^{(f)}(\gamma)$, shown in \cref{circuit1}, is constructed by composing these edge-wise controlled unitaries according to the chosen product formulas. In particular, the first- and second-order constructions are given by
\begin{subequations}\label{Uqw}
\begin{align}
U_{QW}^{(1)}(\gamma) &=
\overset{\rightarrow}{\prod}_{(r,s)} U^{(r,s)}(\gamma) \\
U_{QW}^{(2)}(\gamma) &=
\left[
\overset{\rightarrow}{\prod}_{(r,s)} U^{(r,s)}\!\left(\frac{\gamma}{2}\right)
\right]
\left[
\overset{\leftarrow}{\prod}_{(r,s)} U^{(r,s)}\!\left(\frac{\gamma}{2}\right)
\right]
\end{align}
\end{subequations}
where the arrow $\rightarrow$ denotes a fixed ordering over edges, while $\leftarrow$ denotes the reverse ordering. By setting $\gamma=t/R$ and applying $\left[U_{QW}^{(f)}(t/R)\right]^R$, where $f=1,2$ denotes the first- and second-order product-formula approximations, respectively, we obtain an $R$-step approximation to the controlled time-evolution operator in \cref{controlled_qw_op}, corresponding to a total evolution time $t$. The explicit matrix representation and gate-level decomposition of $U^{(r,s)}(\gamma)$ are discussed in detail in \cref{detailed_implementation}. Moreover, the resulting Trotter error and mitigation strategies are discussed in \cref{Trotter Correction}.

\subsection{Contingency-Aware Centrality Measure}
In this section, we introduce a ranking metric derived from measurement statistics of the quantum circuit in \cref{circuit1}. Before measurement, the circuit produces a joint quantum state of the contingency and walker registers after an evolution time $t$, given by
\begin{equation}
    \ket{\Psi_t} = \Bigl[U_{QW}^{(f)}\left(\frac{t}{R}\right)\Bigr]^R \ket{\Psi_0}
\end{equation}
where $\ket{\Psi_0}$ is defined in \cref{eq:overall_state} and $R$ is the number of Trotter steps and $f$ denotes the order of Trotter-Suzuki formula. Measuring only the walker register in the computational basis $\{\ket{j}\}_{j=1}^n$ yields the probability of finding the walker at node $j$ at time $t$ averaged over contingencies. The contingency-averaged occupation probability is given by $Q_j(t) = \bra{j} \Tr_c\!\big[ \ket{\Psi_t}\!\bra{\Psi_t} \big] \ket{j}$, where $\Tr_c$ is the partial trace over the contingency register. Let the total observation window be $T$, and the number of discrete samples be $S$, with time step $\Delta t=T/S$ and sampling times $t_\tau=\tau\Delta t$ for $\tau=1,\dots,S$. The contingency-aware occupation centrality of node $j$ is then defined as
\begin{equation}\label{eq:cj_metric1}
\Omega_j = \frac{1}{S}\sum_{\tau=1}^{S} Q_j(t_\tau).
\end{equation}
Operationally, the metric $\Omega_j$ is obtained by repeatedly executing the quantum circuit, measuring the walker register in the computational basis at discrete times $t_\tau$, and estimating the probability $Q_j(t_\tau)$ of observing the walker at node $j$. 

\section{Detailed Circuit Implementation} \label{detailed_implementation}
Building on the high-level circuit description introduced in the previous section, here we present the full gate-level realization of the CCTQW. We detail the preparation of the contingency and walker registers, derive the explicit matrix representation of the edge-wise controlled unitary  $U^{(r,s)}(\gamma)$, and describe its decomposition into elementary single- and multi-qubit gates.

\subsection{Block Matrix Form of $U^{(r,s)}(\gamma)$ operator}
In this subsection, we demonstrate how to construct the block matrix representation of the edge-wise controlled unitary operator $U^{(r,s)}(\gamma)$ by utilizing the decomposition of the adjacency matrix. This explicit matrix form is particularly useful for classical simulation of the quantum algorithm and for gaining insight into the internal structure of the walk dynamics. Recall that, for each $(r,s)$ edge we define $H^{(r,s)}=w_{r,s}M^{(r,s)}$. To aid in simplifying powers of $H^{(r,s)}$, let us define a matrix of the form $J^{(r,s)} = v_r v_r^\text{T} + v_s v_s^\text{T}$ which acts as a projector onto the subspace involving nodes $r$ and $s$. It can be shown that the transition matrices satisfy the following algebraic identities
\begin{equation}\label{propety}
    \begin{aligned}
        \left(H^{(r,s)}\right)^{2k} &= w_{r,s}^{2k}J^{(r,s)} \quad , k \geq 1, \\
        \left(H^{(r,s)}\right)^{2k+1} &= w_{r,s}^{2k+1}M^{(r,s)}  \quad , k \geq 0,
    \end{aligned}
\end{equation}
where $k$ is a positive integer. These identities enable us to write the operator $e^{-i\gamma H^{(r,s)}}$ as the matrix
\begin{equation} \label{exp_unitary_closed_form}
    e^{-i\gamma H^{(r,s)}} = \mathbb{I} - i \sin(\gamma w_{r,s}) M^{(r,s)} + \left[\cos(\gamma w_{r,s}) - 1\right] J^{(r,s)}.
\end{equation}

This matrix acts non-trivially only on the subspace spanned by nodes $r$ and $s$ and as the identity on the rest of the Hilbert space. Thus, the block matrix of $ e^{-i\gamma H^{(r,s)}}$ takes the following form
\begin{equation}\label{blox_matrix_of_e}
    \left[ e^{-i\gamma H^{(r,s)}} \right]_{\alpha, \beta} =
    \begin{cases}
        \cos(\gamma w_{r,s}) & \text{if } \alpha = r,\ \beta = r \\
        \cos(\gamma w_{r,s}) & \text{if } \alpha = s,\ \beta = s \\
        -i \sin(\gamma w_{r,s}) & \text{if } \alpha = r,\ \beta = s \\
        -i \sin(\gamma w_{r,s}) & \text{if } \alpha = s,\ \beta = r \\
        1 & \text{if } \alpha = \beta,\ \text{except } r,s \\
        0 & \text{otherwise}
    \end{cases} .
\end{equation}
Hence, using the explicit block matrix form of $e^{-i\gamma H^{(r,s)}}$ in \cref{blox_matrix_of_e}, and substituting it into \cref{Urs}, the block matrix representation of $U^{(r,s)}(\gamma)$ follows directly. 

\subsection{Gate Structure of $U^{(r,s)}(\gamma)$ operator}
For practical realization of CCTQW on quantum hardware, it is essential to efficiently decompose each $U^{(r,s)}(\gamma)$ into a sequence of elementary gates. Insight into this decomposition can be gained from the closed-form expression in \cref{exp_unitary_closed_form}. Each exponential term $e^{-i\gamma H^{(r,s)}}$ operates non-trivially as a rotation only within the two-dimensional subspace spanned by the basis states $|r\rangle$ and $|s\rangle$. This allows us to realize $e^{-i\gamma H^{(r,s)}}$ as a controlled two-level unitary that mixes amplitudes only between $|r\rangle$ and $|s\rangle$. From the block matrix structure shown in \cref{blox_matrix_of_e}, the effective two-level unitary operation on this subspace can be written as
\begin{equation}
F^{(r,s)}=
\bordermatrix{
      & \langle r| & \langle s| \cr
|r\rangle & \cos(\gamma w_{r,s}) & -i\sin(\gamma w_{r,s}) \cr
|s\rangle & -i\sin(\gamma w_{r,s}) & \cos(\gamma w_{r,s}) \cr
}.
\end{equation}
The matrix $F^{(r,s)}$ is structurally equivalent to a single-qubit $R_x$ rotation. Hence, the general construction of $U^{(r,s)}(\gamma)$ involves multi-controlled $R_x$ gates that selectively activate when the corresponding states in the contingency register are $|1\rangle$ and the corresponding input states in the Walker register are either $|r\rangle$ or $|s\rangle$. As an example, we provide explicit gate-level implementations of the $U^{(r,s)}(\gamma)$ related to a CCTQW on the 8-bus power network topology in \cref{sec:8bussystem} below.

\section{Trotter error analysis} \label{Trotter Correction}
The CCTQW operator $U_{QW}^{(f)}(\gamma)$ is implemented as a finite product of edge-wise controlled unitary operators. This introduces an approximation to the ideal continuous-time evolution. Understanding how this error scales with the system size (e.g., number of edges) and how it can be mitigated is therefore essential for the practical implementation of the proposed framework. In general, for a Hamiltonian $H=\sum_{j=1}^{L} H_j$, the number of Trotter steps required to approximate the evolution $e^{-itH}$ with error at most $\varepsilon$ is given by \cite{childs2019faster}
\begin{align}\label{Trotter_step}
\begin{split}
    R_{1} &= \mathcal{O}\!\left( \frac{(t \Lambda L)^2}{\varepsilon} \right) \\
    R_{2k} &= \mathcal{O}\!\left( \frac{(t \Lambda L)^{1 + \tfrac{1}{2k}}}{\varepsilon^{\frac{1}{2k}}} \right)
\end{split}
\end{align}
where $L$ is the number of non-commuting Hamiltonian terms and $\Lambda := \max_j \|H_j\|$ is a spectral-norm bound. Here $R_{1}$ corresponds to the first-order Lie–Trotter formula, while $R_{2k}$ gives the number of steps required for a $(2k)$th-order Suzuki formula. The bounds in \cref{Trotter_step} highlight that the Trotter step is primarily governed by two quantities: the number of terms $L$ and the spectral norm $\Lambda$. In the proposed CCTQW framework, the controlled Hamiltonian evolution is given by \cref{eq:Ue_projector_exp}. From its decomposition, the naive number of Hamiltonian terms is $L = |E|=N$ and the associated spectral-norm parameter is $\Lambda = \max_{(r,s)\in E} \left\| \Pi_1^{(e_{r,s})} \otimes H^{(r,s)} \right\|$. The mitigation strategies discussed in the following subsections are therefore directly aimed at reducing the effective number of terms.

\subsection{Regrouping commuting edge terms to reduce $L$}
\label{sec:sub_regrouping}
A key observation is that edge-local Hamiltonians corresponding to disjoint edges commute. Specifically, if $(r,s)$ and $(r',s')$ share no common node, then $[H^{(r,s)}, H^{(r',s')}] = 0$. This allows the edge set $E$ to be partitioned into commuting groups. Such a partitioning corresponds to the edge-colouring problem, in which each colour class defines a commuting set of edges \cite{cao2019graph}. By Vizing’s theorem, the number of colours and hence the number of commuting groups satisfies $\chi\leq \Delta+1$, where $\Delta$ is the maximum degree of the graph \cite{berge1991short}. Efficient classical algorithms, such as the Misra–Gries edge-colouring algorithm \cite{misra1992constructive}, as well as more recent nearly-linear-time methods \cite{assadi2025vizing}, can construct such a colouring with worst-case runtime $\mathcal{O}(|E|\log\Delta)$.

In our implementation, the edge-coloring is used to organize the controlled edge unitaries $U^{(r,s)}(\gamma)$ according to their commutation structure. We partition the edge set into color classes $\{\Theta_j\}_{j=1}^{\chi}$ such that all edges within each class are mutually disjoint. Accordingly, the corresponding operators $\{U^{(r,s)}(\gamma) : (r,s)\in\Theta_j\}$ are grouped together within the construction of $U_{QW}^{(f)}(\gamma)$. Since the grouping of $U^{(r,s)}(\gamma)$ operators is constructed based on the commutation relations of $H^{(r,s)}$, the same commutation structure is inherited by the resulting sequence of $e^{-i\gamma H^{(r,s)}}$ operators acting on the walker state. This allows the product-formula sequence to be organized into $\chi \leq \Delta+1$ commuting layers.

Typically, power networks are sparse and possess a relatively small maximum nodal degree $\Delta$ compared to the number of buses. Hence, for a power network, the number of commuting groups remains limited even as system size grows. Empirical studies on benchmark transmission networks further support this observation. In particular, for networks from the NESTA archive, it has been shown that the maximum node degree $\Delta$ grows approximately logarithmically with the number of buses \cite{kersulis2018topological}. As a result, $\Delta$ remains bounded by a modest constant for most practical cases, implying that the edge-colouring–based grouping reduces the overall circuit depth, even for large-scale systems.

\subsection{Spectral norm and edge weights}
The number of Trotter steps required to approximate the continuous-time evolution depends not only on the number of non-commuting Hamiltonian terms $L$, but also on the spectral norm bound $\Lambda$, as shown in \cref{Trotter_step}. We now show how edge weights directly influence this quantity and can therefore impact the number of Trotter steps. Recall that, from \cref{eq:Ue_projector_exp}, we can write $\Lambda = \max_{(r,s)\in E} \left\| \Pi_1^{(e_{r,s})} \otimes H^{(r,s)} \right\|$. Using the property $\|A \otimes B\| = \|A\|\,\|B\|$ and noting that $\Pi_1^{(e_{r,s})}$ is an orthogonal projector with $\|\Pi_1^{(e_{r,s})}\| = 1$, we obtain $\Lambda = \max_{(r,s)\in E} \|H^{(r,s)}\|$. Moreover, using the scaling property of the operator norm, $\|H^{(r,s)}\| = |w_{r,s}|\,\|M^{(r,s)}\|$. The matrix $M^{(r,s)}$ acts nontrivially only on the two-dimensional subspace spanned by $\{v_r, v_s\}$. On that subspace, it has the Pauli-$X$ form, while all other entries are zero. The corresponding $2 \times 2$ block has eigenvalues $\pm 1$, and the remaining eigenvalues are all $0$. Since the matrix is real symmetric, its spectral norm equals the largest absolute eigenvalue, and hence $\|M^{(r,s)}\| = 1$. It follows that $\Lambda = \max_{(r,s)\in E} |w_{r,s}|$. This result shows that the spectral-norm parameter governing the Trotter step is determined solely by the maximum edge weight. 

\section{Results} \label{Particular Case}
This section presents the implementation and numerical results of the proposed CCTQW framework on an 8-bus system under $N-2$ contingencies, including a comparison with classical centrality measures.

\subsection{The 8-bus system} \label{sec:8bussystem}

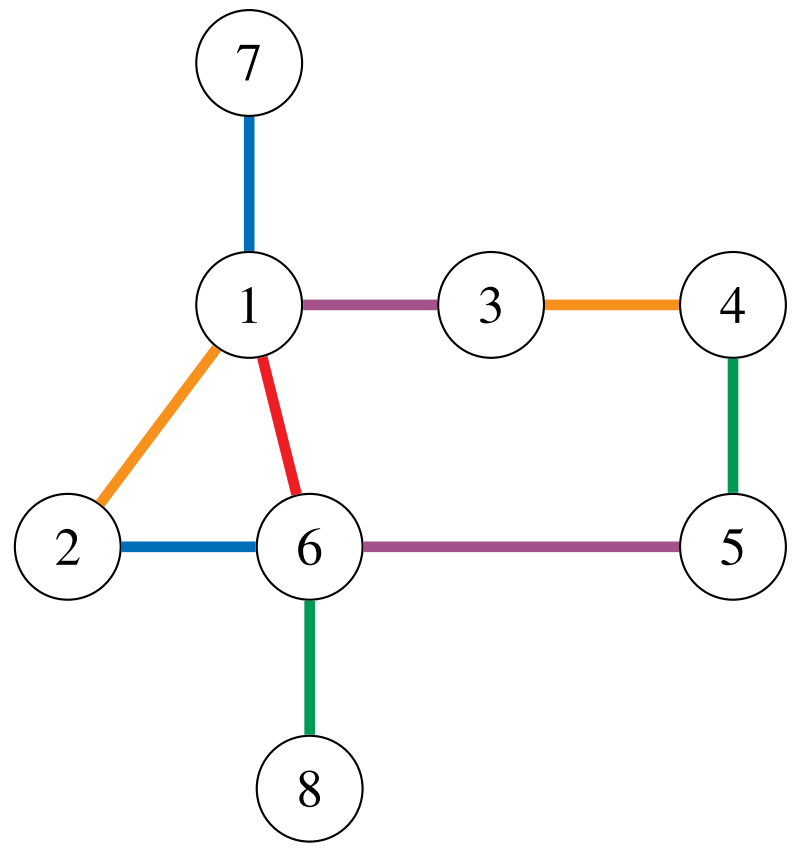
\begin{figure}
    \centering
    \begin{tikzpicture}[scale=0.8]
    \node[draw, circle, minimum size=0.7cm] (bus1) at (0, 0) {1};
    \node[draw, circle, minimum size=0.7cm] (bus2) at (-1.5, -2) {2};
    \node[draw, circle, minimum size=0.7cm] (bus3) at (2, 0) {3};
    \node[draw, circle, minimum size=0.7cm] (bus4) at (4, 0) {4};
    \node[draw, circle, minimum size=0.7cm] (bus5) at (4, -2) {5};
    \node[draw, circle, minimum size=0.7cm] (bus6) at (0.5, -2) {6};
    \node[draw, circle, minimum size=0.7cm] (bus7) at (0, 2) {7};
    \node[draw, circle, minimum size=0.7cm] (bus8) at (0.5, -4) {8};
    \draw[line width=2pt, NavyBlue] (bus1) -- (bus7);
    \draw[line width=2pt, NavyBlue] (bus2) -- (bus6);
    \draw[line width=2pt, Red] (bus1) -- (bus6);
    \draw[line width=2pt, ForestGreen] (bus6) -- (bus8);
    \draw[line width=2pt, ForestGreen] (bus4) -- (bus5);
    \draw[line width=2pt, DarkOrchid] (bus5) -- (bus6);
    \draw[line width=2pt, DarkOrchid] (bus1) -- (bus3);
    \draw[line width=2pt, BurntOrange] (bus3) -- (bus4);
    \draw[line width=2pt, BurntOrange] (bus1) -- (bus2);
    \end{tikzpicture}
    \caption{Graph representation of the 8-bus network. Edges are colored 
    according to their commuting groups used in the edge-coloring based 
    Trotter--Suzuki decomposition: \textcolor{NavyBlue}{Group 1} 
    $\{(1,7),(2,6)\}$, \textcolor{Red}{Group 2} $\{(1,6)\}$, 
    \textcolor{ForestGreen}{Group 3} $\{(6,8),(4,5)\}$, 
    \textcolor{DarkOrchid}{Group 4} $\{(5,6),(1,3)\}$, 
    \textcolor{BurntOrange}{Group 5} $\{(3,4),(1,2)\}$.}
    \label{graph_8_node}
\end{figure}

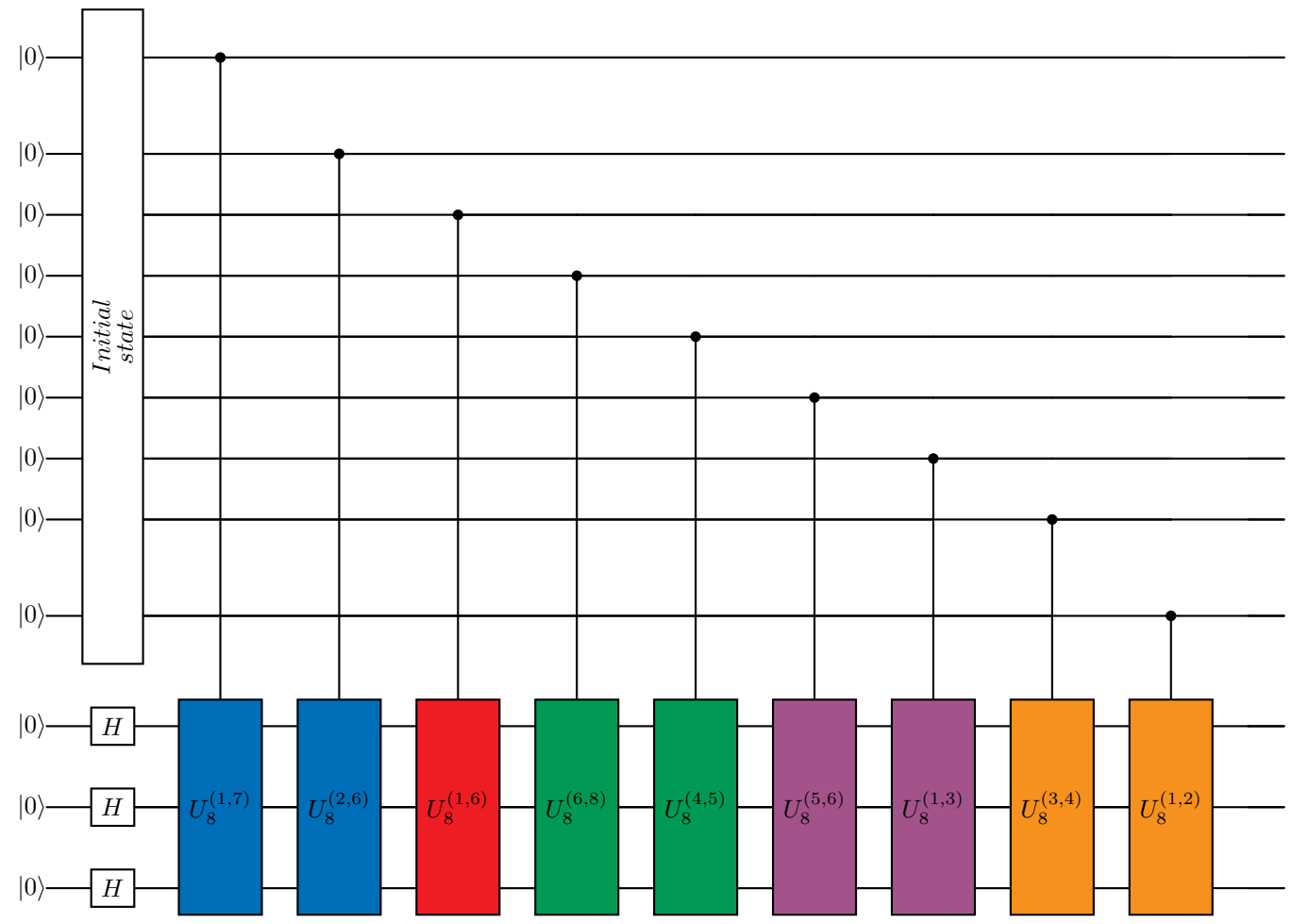
\begin{figure}[t]
    \resizebox{0.5\textwidth}{!}{
    \begin{quantikz}
        \ket{0} & \gate[9]{\rotatebox{90}{\shortstack{$Initial$ \\ $state$}}} 
        & \ctrl{9} & \qw & \qw & \qw & \qw & \qw & \qw & \qw & \qw 
        & & \qw \\
        \ket{0} & & \qw & \ctrl{8} & \qw & \qw & \qw & \qw & \qw & \qw & \qw 
        & & \qw  \\
        \ket{0} & & \qw & \qw & \ctrl{7} & \qw & \qw & \qw & \qw & \qw & \qw 
        & & \qw \\
        \ket{0} & & \qw & \qw & \qw & \ctrl{6} & \qw & \qw & \qw & \qw & \qw 
        & & \qw \\
        \ket{0} & & \qw & \qw & \qw & \qw & \ctrl{5} & \qw & \qw & \qw & \qw 
        & & \qw  \\
        \ket{0} & & \qw & \qw & \qw & \qw & \qw & \ctrl{4} & \qw & \qw & \qw 
        & & \qw  \\
        \ket{0} & & \qw & \qw & \qw & \qw & \qw & \qw & \ctrl{3} & \qw & \qw 
        & & \qw \\
        \ket{0} & & \qw & \qw & \qw & \qw & \qw & \qw & \qw & \ctrl{2} & \qw 
        & & \qw  \\
        \ket{0} & & \qw & \qw & \qw & \qw & \qw & \qw & \qw & \qw & \ctrl{1} 
        & & \qw \\ 
        \ket{0} & \gate{H} 
        & \gate[3, style={fill=NavyBlue}]{U^{(1,7)}_8} 
        & \gate[3, style={fill=NavyBlue}]{U^{(2,6)}_8} 
        & \gate[3, style={fill=Red}]{U^{(1,6)}_8} 
        & \gate[3, style={fill=ForestGreen}]{U^{(6,8)}_8} 
        & \gate[3, style={fill=ForestGreen}]{U^{(4,5)}_8} 
        & \gate[3, style={fill=DarkOrchid}]{U^{(5,6)}_8} 
        & \gate[3, style={fill=DarkOrchid}]{U^{(1,3)}_8} 
        & \gate[3, style={fill=BurntOrange}]{U^{(3,4)}_8} 
        & \gate[3, style={fill=BurntOrange}]{U^{(1,2)}_8} 
        & & \qw \\
        \ket{0} & \gate{H} & \qw & \qw & \qw & \qw & \qw & \qw & \qw & \qw & \qw 
        & & \qw \\
        \ket{0} & \gate{H} & \qw & \qw & \qw & \qw & \qw & \qw & \qw & \qw & \qw 
        & & \qw \\
    \end{quantikz}
    } 
    \caption{Internal structure of the first-order CCTQW operator $U_{QW}^{(1)}(\gamma)$ for the 8-bus system. The controlled edge-wise unitaries are applied in a fixed ordering, corresponding to the forward pass. Gate colors indicate commuting groups, consistent with \cref{graph_8_node}.}  
\label{circuit_8_node}
\end{figure}

The associated graph representation of the 8-bus system is shown in \cref{graph_8_node}. The system consists of $N=9$ transmission lines, and we consider contingency scenario $x=2$, corresponding to all double-line outage configurations.
The adjacency matrix of the weighted graph representing the 8-bus system is denoted by $\mathcal{A}^{(8)}_w=[w_{rs}]$. The weights are defined for unordered pairs $r<s$ such that $w_{rs}=w_{sr}$ and $w_{rr}=0$. The nonzero edge weights are chosen as $w_{13}=w_{56}=0.12$, $w_{34}=w_{45}=0.11$, $w_{16}=0.20$, $w_{12}=w_{26}=0.09$, and $w_{17}=w_{68}=0.08$ , while all remaining entries are zero. These values are selected to represent a stronger meshed backbone $1-6-5-4-3-1$ and a weaker peripheral corridor $7-1-2-6-8$. For our quantum approach, we normalize the weights so that $\sum_{\{ r,s\}} w_{rs}=1$. The corresponding quantum circuit diagram which shows the internal structure of the CCTQW on the 8-bus system is given in \cref{circuit_8_node}. 

\begin{figure}
\centering
$U^{(1,3)}_8(\gamma)$ = 
\begin{quantikz}[row sep=0.5cm, column sep=0.5cm]
\lstick{$c_7$} & \qw & \ctrl{3} & \qw & \qw\\
\lstick{$q_0$} & \gate{X} & \ctrl{2} & \gate{X} \\
\lstick{$q_1$} & \qw      & \gate{R_x(\gamma)} & \qw \\
\lstick{$q_2$} & \gate{X} & \ctrl{-1} & \gate{X}
\end{quantikz}
\caption{Gate-level realization of the unitary operator $U^{(1,3)}_8(\gamma)$ corresponding to the edge $(1,3)$ in the 8-bus system. The qubits $q_i \in \{0,1,2\}$ belong to the walker register and the qubit $c_7$, which belongs to the contingency register, corresponds to the edge $(1,3)$.}
\label{fig:u_gate_decom}
\end{figure}
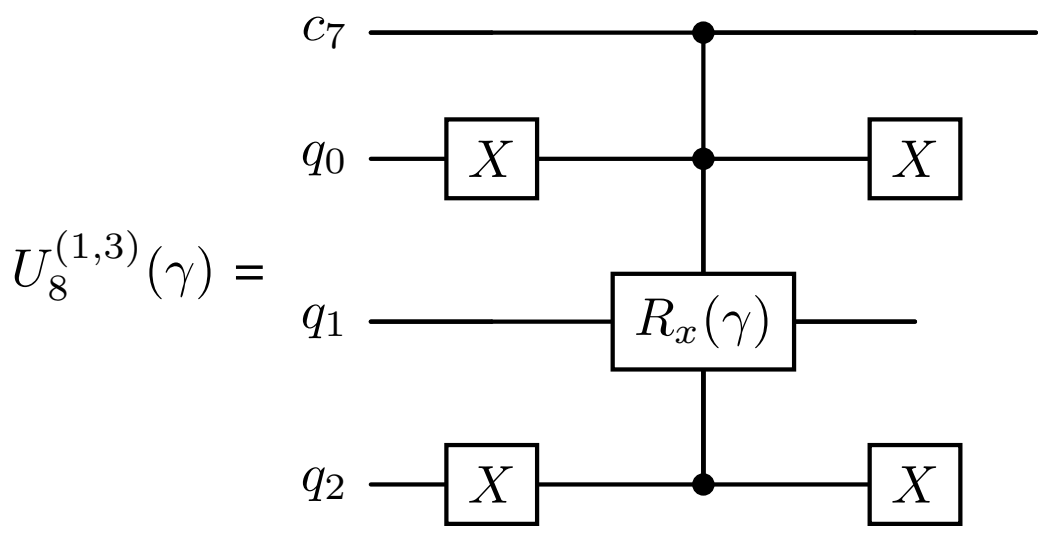

For illustration, \cref{fig:u_gate_decom} shows the gate decomposition of the unitary corresponding to edge $(1,3)$. The unitaries for the remaining edges are constructed using a single $R_x$ rotation together with a combination of $X$ and CNOT gates.
The evolved joint state is given by $ |\tilde{\Psi_t}\rangle = \left[U_{QW}^{(1)}\left(\frac{t}{R}\right) \right]^R  |\tilde{D}^9_x\rangle \otimes |\tilde\phi_0\rangle$ where the $|\tilde{D}^9_x\rangle$ is the Dicke state after applying the multi controlled Pauli-X gate as described in \cref{sec:state_prep} and $|\tilde\phi_0\rangle$ initial states are defined using \cref{dicke state,initial_walker} forming the intial state \cref{eq:overall_state}. In this work, we employ the first-order Trotter--Suzuki formula. The numerical results are obtained using algebraic simulations, while equivalent Qiskit circuit simulations reproduce the same results.

\subsection{Numerical Error Analysis}
\begin{figure}
\centering
\includegraphics[width=\linewidth]{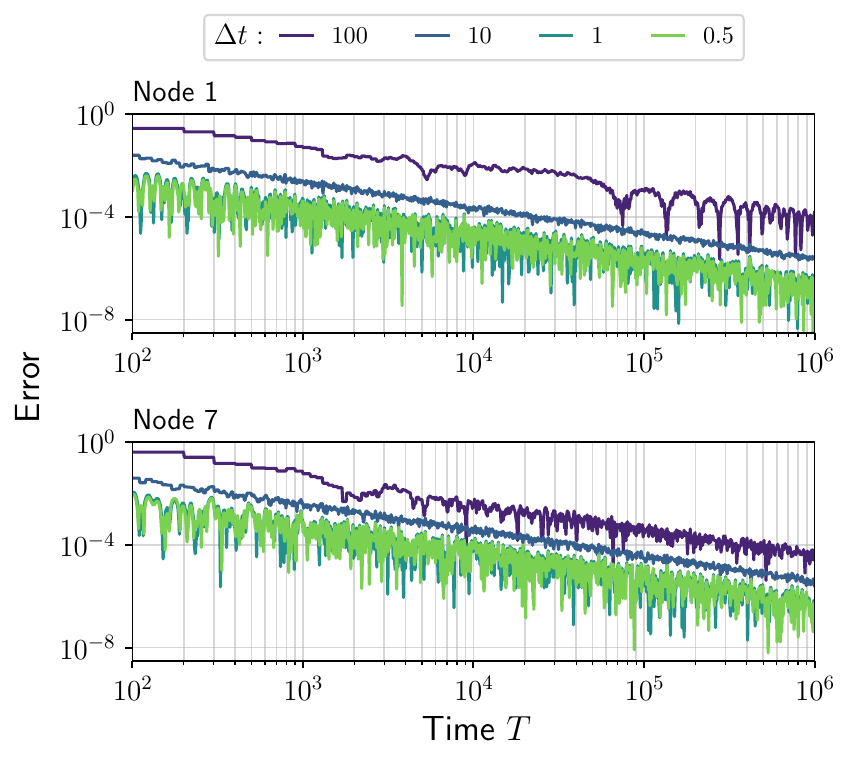}
\caption{ Relative error $|\Omega_j(T,\Delta t,\infty)-\Omega_j^{\infty}|/\Omega_j^{\infty}$ for the weighted 8-bus system under $N$-2 contingencies using the exact time average $\Omega_j^{\infty}$ as a function of the total time horizon $T$ and four different sampling steps $\Delta t$. 
Both axes are logarithmically scaled. We show backbone node (Node~1, top) and a peripheral node (Node~7, bottom). All four sampling steps $\Delta t$ show a decay with $T$ and reach comparable accuracy at $T=10^6$. The fine resolutions $\Delta t \leq 1$ nearly coincide for the full time horizon. Visible oscillations come from possible cancellations between eigenvalue-difference frequencies, so isolated dips should not be understood as improved convergence.}
\label{fig:finte_time_error}
\end{figure}

For the numerical analysis, we explicitly examine the dependence of the contingency-averaged occupation centrality on the finite observation horizon $T$, sampling interval $\Delta t$, and number of Trotter steps $R$. Accordingly, we denote the finite numerical estimate as $\Omega_j(T,\Delta t,R)$. This notation represents the same metric $\Omega_j$ defined in \cref{eq:cj_metric1}, while making its dependence on the numerical parameters explicit. We first define the asymptotic contingency-averaged occupation centrality, which serves as the reference for the numerical analysis. For the $k$-th contingency-induced graph, let $H^{(k)}=\sum_{\lambda}\lambda\,\Pi_{\lambda}^{(k)}$ denote the spectral decomposition of its Hamiltonian, where $\Pi_{\lambda}^{(k)}$ is the projector onto the eigenspace associated with eigenvalue $\lambda$. Using \cref{centrality_measure}, the exact infinite-time occupation centrality at node $j$ of the $k$-th contingency-induced graph is $\tilde{C}_{j,\mathrm{exact}}^{(k)}=\sum_{\lambda}|\langle j|\Pi_{\lambda}^{(k)}|\tilde\phi_0\rangle|^2$. Therefore, we define the asymptotic contingency-averaged occupation centrality as
\begin{equation}
  \Omega_j^{\infty} =
\frac{1}{\binom{9}{2}}
\sum_{k=1}^{\binom{9}{2}}
\tilde{C}_{j,\mathrm{exact}}^{(k)}.
\end{equation}
Next we examine the effect of the finite observation horizon  $T$ and the sampling interval $\Delta t$, independently of the Trotter approximation. For this purpose, the occupation probabilities of the $k$-th graph are evaluated using the exact propagator, $P_j^{(k)}(t_\tau)=|\langle j|e^{-iH^{(k)}t_\tau}|\tilde\phi_0\rangle|^2$.
We denote the corresponding finite-time contingency-averaged occupation centrality by

\begin{equation}
\Omega_j(T,\Delta t,\infty) = \frac{1}{\binom{9}{2}}
\sum_{k=1}^{\binom{9}{2}}
\frac{1}{S}
\sum_{\tau=1}^{S}
P_j^{(k)}(t_\tau) \ \text{and} \
S=\frac{T}{\Delta t} 
\end{equation}
where $t_\tau=\tau\Delta t$ for $\tau=1,\dots,S$. Here, \(R=\infty\) indicates that the exact propagator is used and no Trotter approximation is introduced. \cref{fig:finte_time_error} shows the relative error as a function of the observation horizon $T$ for different sampling intervals $\Delta t$. For clarity, the figure presents the results for Nodes~1 and~7 as representative examples. As shown in \cref{fig:finte_time_error}, the relative error generally decreases with increasing observation horizon $T$ for all considered sampling intervals. In particular, the results obtained with $\Delta t=0.5$ and $\Delta t=1$ are nearly indistinguishable over the considered range of $T$. This indicates that reducing the sampling interval below $\Delta t=1$ provides no appreciable improvement for the present analysis. Therefore, we set $\Delta t=1$ for the subsequent numerical investigations.

\begin{figure}
\centering
\includegraphics[width=\linewidth]{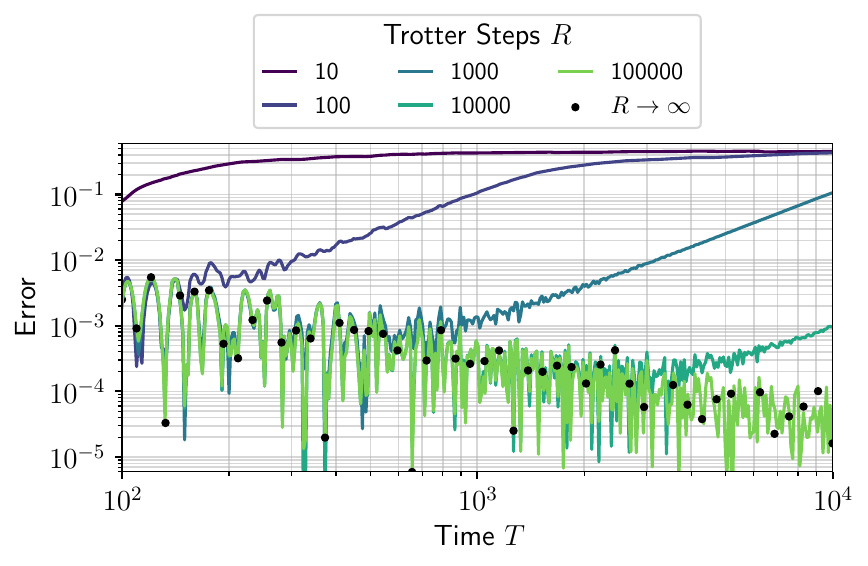}
\caption{ Relative error $|\Omega_j(T,1,R)-\Omega_j^{\infty}|/\Omega_j^{\infty}$ for the weighted 8-bus system under $N$-2 contingencies using the exact time average $\Omega_j^{\infty}$ as a function of the total time horizon $T$ and at fixed $\Delta t=1$. The different curves show different finite Trotter steps $R$. The dots indiciate results for $R \rightarrow \infty$, implying that no Trotter error is introduced.}
    \label{fig:trotter_error}
\end{figure}

We next investigate the influence of the finite observation horizon $T$ and the number of Trotter steps $R$. We denote the resulting finite-time, Trotterized estimate by $\Omega_j(T,1,R)$. \cref{fig:trotter_error} shows the relative error for Node 1 as a function of the observation horizon $T$ for different values of $R$. The results show that both $T$ and $R$ influence the accuracy of the estimated occupation centrality. In particular, as $R$ increases, the finite-$R$ results approach the $R\rightarrow\infty$ reference, demonstrating the convergence of the Trotterized evolution toward the exact time evolution. 

\begin{figure}
\centering
\includegraphics[width=\linewidth]{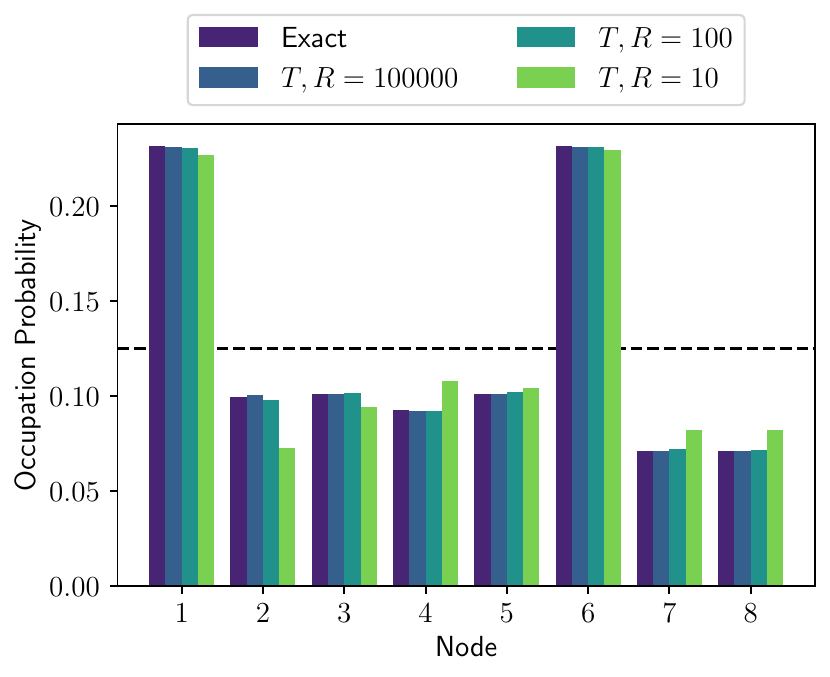}
\caption{\justifying Contingency-averaged occupation probability $\Omega_j$ for all eight nodes of the weighted 8-bus system under $N - 2$ contingencies. The most left bar shows the exact asymptotic value $\Omega_j^{\infty}$. The following bars are associated to a finite observation window $T$ and step count $R$, whereby we choose $T=R \in \{10,100,100000\}$ for $\Delta t = 1$. The dashed line marks the uniform initial probability $1/8$. Notably,
even low $T, R$ suffice to achieve a node ranking that matches the exact node ranking qualitatively.} 
\label{fig: ranking comparison}
\end{figure}

The primary objective of the proposed metric $\Omega_j$ in \cref{eq:cj_metric1} is to rank the nodes according to their contingency-averaged importance. We therefore examine whether the node ranking can already be recovered for finite values of $T$ and $R$. \cref{fig: ranking comparison} shows $\Omega_j(T,\Delta t=1, R)$ for all eight nodes of the weighted 8-bus system under $N-2$ contingencies using $T=R\in\{10,100,100000\}$. For the asymptotic reference, the node ordering is $ 1=6>3=5>2>4>7=8$. According to \cref{fig: ranking comparison}, even for relatively small $T$ and $R$, the dominant structure of the ranking is already visible, with Nodes $1$ and $6$ having the largest occupation centralities and Nodes $7$ and $8$ remaining among the least important nodes. The ordering of the intermediate nodes is more sensitive to finite-$T$ and finite-$R$ effects. As $T$ and $R$ increase, the finite estimates approach the asymptotic values and the corresponding node ordering converges toward the exact ranking. Importantly, even for $T=R=100$, the dominant ranking structure is already captured, with Nodes $1$ and $6$ identified as the most central and Nodes $7$ and $8$ among the least central. This demonstrates that a meaningful node ranking can be obtained using relatively small values of $T$ and $R$, without requiring full numerical convergence of $\Omega_j$. This is particularly relevant for the proposed application, since the objective is to identify the relative importance of the nodes rather than to reproduce the asymptotic centrality values with arbitrarily high precision. 

\subsection{Comparison to Classical Centrality}
\label{classical_centrality}
To provide a qualitative comparison with classical centrality-based contingency screening approaches, we evaluate the proposed CCTQW-based framework against betweenness centrality, closeness centrality, and current-flow betweenness centrality on the weighted 8-bus system under $N-2$ contingencies. We adopt the approach in \cite{coelho2018new} and extend the evaluation of these measures from individual network configurations to an ensemble of contingency-induced graphs. The 8-bus system consists of $N=9$ transmission lines, yielding $\binom{9}{2}=36$ double-line outage configurations. Excluding configurations that result in network islanding leaves 14 connected configurations. Let $C_{j,\mathrm{classical}}^{(k)}$ denote the centrality of node $j$ in configuration $k$. The aggregated centrality is calculated as $\tilde{C}_{j,\mathrm{classical}} = \sum_{k=1}^{14} C_{j,\mathrm{classical}}^{(k)}$. The resulting values are then normalized across all nodes as $\tilde{C}_{j,\mathrm{classical}} /
\sum_{i=1}^{8}\tilde{C}_{i,\mathrm{classical}}$. This procedure is applied to each of the three classical centrality measures. 

\begin{figure}[ht]
    \centering
    \includegraphics[width=1\linewidth]{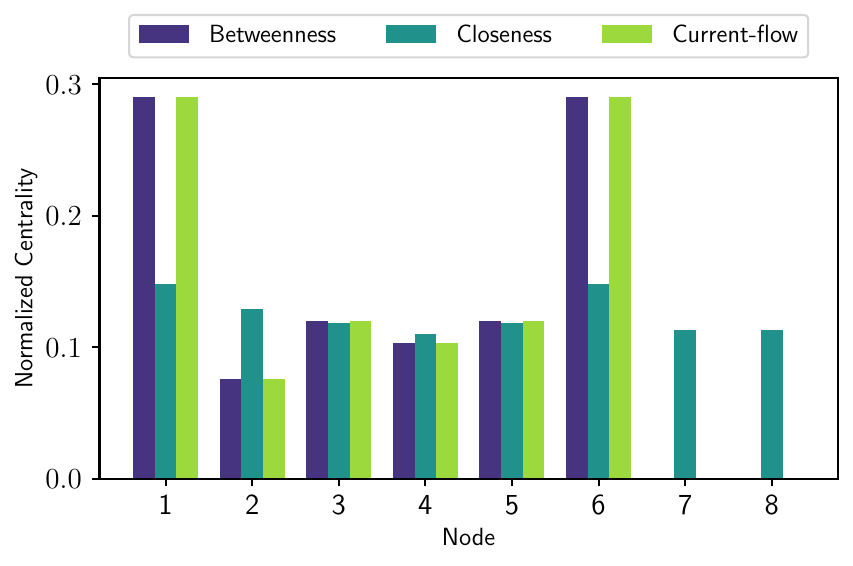}
    \caption{Normalized classical centrality measures for the
    weighted 8-bus system under $N-2$ contingencies,
    aggregated over the connected contingency-induced graphs.}
    \label{centrality measures for 8-bus}
\end{figure}

As illustrated in \cref{centrality measures for 8-bus}, the three classical centrality measures yield different node rankings under $N-2$ contingencies. Betweenness centrality and current-flow betweenness centrality assign the highest values to Nodes 1 and 6, while closeness centrality produces a more uniform ranking with comparatively smaller differences among the nodes. In particular, betweenness and current-flow betweenness centralities assign zero values to peripheral Nodes 7 and 8, whereas closeness centrality yields nonzero values for these nodes.

The classical results show qualitative agreement with the proposed CCTQW-based metric, particularly in identifying Nodes 1 and 6 as the most central. However, while betweenness and current-flow betweenness centralities assign zero values to peripheral nodes, the CCTQW-based metric assigns nonzero occupation probabilities to all nodes, allowing them to be ranked according to their contingency-averaged occupation centrality $\Omega_j$. Overall, the comparison demonstrates that the proposed CCTQW framework produces node rankings that are qualitatively consistent with established classical centrality measures.

\section{Discussion} \label{Discussion}

The main computational motivation of the proposed framework is that the number of contingency scenarios grows combinatorially as $\binom{N}{x}$, where $N$ is the number of transmission lines and $x$ is the number of simultaneous outages. In a classical approach, centrality measures must be evaluated separately for each of the $\binom{N}{x}$ contingency-induced subgraphs. In contrast, the proposed CCTQW framework encodes all $N-x$ contingency configurations in a quantum superposition in the form of a Dicke state and evolves the quantum walker conditionally on the encoded network topology. As a consequence, for a fixed time, the resulting state contains branches corresponding to the walk dynamics on all subgraphs in the chosen contingency class.

The node-ranking metric $\Omega_j$ is obtained from repeated measurements and reflects aggregated statistical properties of the contingency ensemble rather than individual contingency outcomes. Estimating it from repeated measurements does not by itself establish a quantum speedup over classical sampling of the same ensemble. Nevertheless, the proposed CCTQW framework could provide a foundation for future Grover-type search algorithms aimed at identifying relevant contingency configurations. Such an extension would require defining a suitable configuration-dependent importance score and evaluating it coherently to construct a marking oracle for configurations represented by computational basis states with Hamming weight $N-x$.

We now discuss the quantum resources required to realize this framework. The required number of qubits scales as $N + \lceil \log_2 n \rceil$ where $N$ qubits are used for the contingency register and $\lceil \log_2 n \rceil$ qubits are used for the walker register. The circuit depth is determined by three main components: the Dicke-state preparation for the contingency register, the preparation of the initial walker state, and the implementation of the controlled Hamiltonian evolution. For product-formula approach, the depth scales with the number of Trotter steps $R$, the number of controlled edge-local operators, and the gate decomposition cost of each $U^{(r,s)}(\gamma)$. The number of Trotter steps can be significantly reduced by regrouping commuting edge terms. Since power networks are inherently sparse, the number of edge-local Hamiltonian terms remains relatively small compared to Hamiltonians typically encountered in quantum chemistry or many-body simulations. Consequently, the overall product-formula implementation is more structured and potentially less demanding in circuit depth. The gate decomposition cost of each $U^{(r,s)}(\gamma)$ depends on the size of the walker register, as the operator selectively applies an $R_x$ rotation on the subspace spanned by the states $\{|r\rangle, |s\rangle\}$ while leaving all other basis states unchanged.

The measurement cost depends on the desired statistical accuracy of the estimated probabilities $Q_j(t_\tau)$. Since $\Omega_j$ is obtained by averaging over $S$ sampling times, the circuit must be executed repeatedly for each time point to estimate the walker occupation probabilities. Thus, although the CCTQW avoids treating every contingency subgraph separately, it introduces a sampling overhead. The total sampling cost depends on the target precision of the estimated node probabilities. Whether this sampling overhead is favorable compared to classical centrality evaluation on each of the $\binom{N}{x}$ subgraphs separately remains an open question and is left for future analysis.

\section{Conclusion}
This paper presented a Controlled Continuous-Time Quantum Walk framework for power-grid analysis under $N-x$ contingencies. The proposed framework encodes all $\binom{N}{x}$ contingency-induced subgraphs in a quantum superposition, enabling a single coherent QW evolution over the entire contingency ensemble. The key contribution of this paper is the formulation and circuit-level realization of this controlled quantum-walk methodology. To demonstrate the applicability of the proposed framework to power-grid analysis under $N-x$ contingencies, a contingency-averaged occupation centrality measure was defined to rank nodes according to their structural importance. Numerical results on an 8-bus system under $N-2$ contingencies showed qualitative agreement with established classical centrality measures, particularly in identifying the most central nodes. Importantly, the dominant node-ranking structure can be recovered using relatively small values of $T$ and $R$, without requiring full numerical convergence of the centrality values. This suggests that meaningful results can be obtained with fewer Trotter steps and shorter observation time horizons, potentially reducing the circuit depth and overall computational requirements of the proposed framework. These findings provide a foundation for further investigations into quantum-walk-based approaches to power-system security assessment.

Nevertheless, several limitations remain. First, the current validation is restricted to topology-based analysis on a small test system, and the relationship between the proposed metric and physical security indicators (e.g., overloads or voltage violations) has not yet been established. Second, rigorous computational advantage over classical approaches is still to be demonstrated. Third, the circuit depth required for accurate Hamiltonian simulation may limit applicability on near-term quantum hardware.

Future work should focus on integrating power-flow-based metrics, validating performance on large-scale systems, and establishing clear comparisons with classical methods for critical component identification under contingencies. Additionally, exploring alternative quantum implementations with reduced depth and resource requirements will be essential for practical scalability.

Finally, the proposed framework can be viewed as Grover-inspired, in the sense that a superposition of computational basis states provides a representation of the contingency search space. In a full Grover-type search, a marking oracle is required to identify relevant contingency configurations via a phase flip. Defining such an oracle is non-trivial and would require a coherent and reversible implementation of an appropriate configuration-dependent importance criterion. In the present work, the centrality measure is defined directly from measurement statistics of the CCTQW, which does not by itself imply a quantum advantage over classical sampling. Nevertheless, the resulting node-ranking score could provide a starting point for developing suitable configuration-dependent objective functions for such an oracle. Identifying and systematically analyzing suitable walk-based scores for this purpose remains an important direction for future work.

\section*{Acknowledgments}
The paper was written as part of the project ``Quantum-based Energy Grids (QuGrids)”, which is receiving funding from the programme ``Profilbildung 2022”, an initiative of the Ministry of Culture and Science of the State of North Rhine-Westphalia.

\bibliographystyle{IEEEtran}
\bibliography{citations}

@PREAMBLE{
 "\providecommand{\noopsort}[1]{}" 
 # "\providecommand{\singleletter}[1]{#1}%" 
}

@inproceedings{halappanavar2012towards,
  title={Towards efficient nx contingency selection using group betweenness centrality},
  author={Halappanavar, Mahantesh and Chen, Yousu and Adolf, Robert and Haglin, David and Huang, Zhenyu and Rice, Mark},
  booktitle={2012 SC Companion: High Performance Computing, Networking Storage and Analysis},
  pages={273--282},
  year={2012},
  organization={IEEE}
}

@article{coelho2018new,
  title={A new approach for contingency analysis based on centrality measures},
  author={Coelho, Elizandra Pereira Roque and Paiva, Marcia Helena Moreira and Segatto, Marcelo Eduardo Vieira and Caporossi, Gilles},
  journal={IEEE Systems Journal},
  volume={13},
  number={2},
  pages={1915--1923},
  year={2018},
  publisher={IEEE}
}

@article{srivastava2018graph,
  title={Graph-theoretic algorithms for cyber-physical vulnerability analysis of power grid with incomplete information},
  author={Srivastava, Anurag K and Ernster, Timothy A and Liu, Ren and Krishnan, Vignesh G},
  journal={Journal of Modern Power Systems and Clean Energy},
  volume={6},
  number={5},
  pages={887--899},
  year={2018},
  publisher={SGEPRI}
}

@article{narimani2021generalized,
  title={Generalized contingency analysis based on graph theory and line outage distribution factor},
  author={Narimani, Mohammad Rasoul and Huang, Hao and Umunnakwe, Amarachi and Mao, Zeyu and Sahu, Abhijeet and Zonouz, Saman and Davis, Katherine},
  journal={IEEE Systems Journal},
  volume={16},
  number={1},
  pages={626--636},
  year={2021},
  publisher={IEEE}
}

@article{kempe2003quantum,
  title={Quantum random walks: an introductory overview},
  author={Kempe, Julia},
  journal={Contemporary Physics},
  volume={44},
  number={4},
  pages={307--327},
  year={2003},
  publisher={Taylor \& Francis}
}

@article{shenvi2003quantum,
  title={Quantum random-walk search algorithm},
  author={Shenvi, Neil and Kempe, Julia and Whaley, K Birgitta},
  journal={Physical Review A},
  volume={67},
  number={5},
  pages={052307},
  year={2003},
  publisher={APS}
}

@article{kadian2021quantum,
  title={Quantum walk and its application domains: A systematic review},
  author={Kadian, Karuna and Garhwal, Sunita and Kumar, Ajay},
  journal={Computer Science Review},
  volume={41},
  pages={100419},
  year={2021},
  publisher={Elsevier}
}

@article{childs2009universal,
  title={Universal computation by quantum walk},
  author={Childs, Andrew M},
  journal={Physical review letters},
  volume={102},
  number={18},
  pages={180501},
  year={2009},
  publisher={APS}
}

@article{wang2013physical,
  title={Physical implementation of quantum walks},
  author={Wang, Jingbo and Manouchehri, Kia},
  journal={Heidelberg, Springer Berlin},
  volume={10},
  pages={978--3},
  year={2013},
  publisher={Springer}
}

@article{boito2023ranking,
  title={Ranking nodes in directed networks via continuous-time quantum walks},
  author={Boito, Paola and Grena, Roberto},
  journal={Quantum Information Processing},
  volume={22},
  number={6},
  pages={246},
  year={2023},
  publisher={Springer}
}

@inproceedings{rossi2014node,
  title={Node centrality for continuous-time quantum walks},
  author={Rossi, Luca and Torsello, Andrea and Hancock, Edwin R},
  booktitle={Structural, Syntactic, and Statistical Pattern Recognition: Joint IAPR International Workshop, S+ SSPR 2014, Joensuu, Finland, August 20-22, 2014. Proceedings},
  pages={103--112},
  year={2014},
  organization={Springer}
}

@article{wang2022continuous,
  title={Continuous-time quantum walk based centrality testing on weighted graphs},
  author={Wang, Yang and Xue, Shichuan and Wu, Junjie and Xu, Ping},
  journal={Scientific Reports},
  volume={12},
  number={1},
  pages={6001},
  year={2022},
  publisher={Nature Publishing Group UK London}
}

@inproceedings{bartschi2022short,
  title={Short-depth circuits for Dicke state preparation},
  author={B{\"a}rtschi, Andreas and Eidenbenz, Stephan},
  booktitle={2022 IEEE International Conference on Quantum Computing and Engineering (QCE)},
  pages={87--96},
  year={2022},
  organization={IEEE}
}

@article{dicke1954coherence,
  title = {Coherence in Spontaneous Radiation Processes},
  author = {Dicke, R. H.},
  journal = "\href{https://link.aps.org/doi/10.1103/PhysRev.93.99}{Phys. Rev.}",
  volume = {93},
  issue = {1},
  pages = {99--110},
  numpages = {0},
  year = {1954},
  month = {Jan},
  publisher = {American Physical Society},
  doi = {10.1103/PhysRev.93.99},
  url = {https://link.aps.org/doi/10.1103/PhysRev.93.99}
}

@article{childs2019faster,
  title={Faster quantum simulation by randomization},
  author={Childs, Andrew M and Ostrander, Aaron and Su, Yuan},
  journal={Quantum},
  volume={3},
  pages={182},
  year={2019},
  publisher={Verein zur F{\"o}rderung des Open Access Publizierens in den Quantenwissenschaften}
}

@article{zhang2023low,
  title={Low-depth Hamiltonian simulation by an adaptive product formula},
  author={Zhang, Zi-Jian and Sun, Jinzhao and Yuan, Xiao and Yung, Man-Hong},
  journal={Physical Review Letters},
  volume={130},
  number={4},
  pages={040601},
  year={2023},
  publisher={APS}
}

@article{campbell2019random,
  title={A random compiler for fast Hamiltonian simulation},
  author={Campbell, Earl},
  journal={Physical Review Letters},
  volume={123},
  number={7},
  pages={070503},
  year={2019},
  publisher={American Physical Society},
  doi={10.1103/PhysRevLett.123.070503}
}

@article{cao2019graph,
  title={Graph edge coloring: a survey},
  author={Cao, Yan and Chen, Guantao and Jing, Guangming and Stiebitz, Michael and Toft, Bjarne},
  journal={Graphs and Combinatorics},
  volume={35},
  number={1},
  pages={33--66},
  year={2019},
  publisher={Springer}
}

@article{berge1991short,
  title={A short proof for a generalization of Vizing's theorem},
  author={Berge, Claude and Fournier, Jean Claude},
  journal={Journal of graph theory},
  volume={15},
  number={3},
  pages={333--336},
  year={1991},
  publisher={Wiley Online Library}
}

@article{misra1992constructive,
  title={A constructive proof of Vizing's theorem},
  author={Misra, Jayadev and Gries, David},
  journal={Information Processing Letters},
  volume={41},
  number={3},
  pages={131--133},
  year={1992},
  publisher={Elsevier}
}

@inproceedings{assadi2025vizing,
  title={Vizing’s theorem in near-linear time},
  author={Assadi, Sepehr and Behnezhad, Soheil and Bhattacharya, Sayan and Costa, Mart{\'\i}n and Solomon, Shay and Zhang, Tianyi},
  booktitle={Proceedings of the 57th Annual ACM Symposium on Theory of Computing},
  pages={24--35},
  year={2025}
}

@inproceedings{kersulis2018topological,
  title={Topological graph metrics for detecting grid anomalies and improving algorithms},
  author={Kersulis, Jonas A and Hiskens, Ian A and Coffrin, Carleton and Molzahn, Daniel K},
  booktitle={2018 Power Systems Computation Conference (PSCC)},
  pages={1--7},
  year={2018},
  organization={IEEE}
}

@article{chen2024implementation,
  title={Implementation of a continuous-time quantum walk on a sparse graph},
  author={Chen, Zhaoyang and Li, Guanzhong and Li, Lvzhou},
  journal={Physical Review A},
  volume={110},
  number={5},
  pages={052215},
  year={2024},
  publisher={APS}
}

@article{loke2017efficient,
  title={Efficient quantum circuits for continuous-time quantum walks on composite graphs},
  author={Loke, Thomas and Wang, Jingbo B},
  journal={Journal of Physics A: Mathematical and Theoretical},
  volume={50},
  number={5},
  pages={055303},
  year={2017},
  publisher={IOP Publishing}
}

@article{portugal2025efficient,
  title={Efficient Circuit Implementations of Continuous-Time Quantum Walks for Quantum Search},
  author={Portugal, Renato and Moqadam, Jalil Khatibi},
  journal={Entropy},
  volume={27},
  number={5},
  pages={454},
  year={2025},
  publisher={MDPI}
}

@article{soklakov2006efficient,
  title={Efficient state preparation for a register of quantum bits},
  author={Soklakov, Andrei N and Schack, R{\"u}diger},
  journal={Physical Review A—Atomic, Molecular, and Optical Physics},
  volume={73},
  number={1},
  pages={012307},
  year={2006},
  publisher={APS}
}

@article{mottonen2004transformation,
  title={Transformation of quantum states using uniformly controlled rotations},
  author={Mottonen, Mikko and Vartiainen, Juha J and Bergholm, Ville and Salomaa, Martti M},
  journal={arXiv preprint quant-ph/0407010},
  year={2004}
}

@techreport{ENTSOE2025GridIncident,
  author      = {{ENTSO-E}},
  title       = {Final Report on the Grid Incident in South-East Europe on 21 June 2024},
  institution = {European Network of Transmission System Operators for Electricity (ENTSO-E)},
  address     = {Brussels, Belgium},
  month       = feb,
  year        = {2025},
  url         = {https://www.entsoe.eu/news/2025/02/25/entso-e-publishes-the-final-report-on-the-grid-incident-in-south-east-europe/}
}

\end{document}